# Graphene for Antenna Applications: Opportunities and Challenges from Microwaves to THz

Julien Perruisseau-Carrier
Adaptive MicroNano Wave Systems, LEMA/Nanolab
Ecole Polytechnique Fédérale de Lausanne (EPFL), Switzerland
julien.perruisseau-carrier@epfl.ch

*Abstract*— **The use of graphene for antennas and other electromagnetic passives could bring significant benefit such as extreme miniaturization, monolithic integration with graphene RF nanoelectronics, efficient dynamic tuning, and even transparency and mechanical flexibility. Though recently different related theoretical works have been presented, relatively few applications have been proposed and realistically assessed. In this invited talk we will briefly review the main properties of graphene and the state of the art in its theoretical and experimental characterization. Then, we will discuss a number of potential antenna applications from microwave to THz, providing in each case a critical assessment of the benefits, limitations, and remaining issues towards actual real-life implementations. Here we provide a brief overview of different devices and associated developments in our group discussed in the talk, including graphene antennas and reflectarrays at microwave and THz, plasmonic switches, isotropic and anisotropic meta-surfaces, or graphene RF-NEMS.**

## I. Introduction

Graphene has recently attracted tremendous interest in various research areas due to its exceptional electrical and mechanical properties [1]. Its unique band structure is notably being exploited in nanoelectronics to achieve novel high-speed devices such as field effect transistors and frequency multipliers [2-4]. Nevertheless the application of graphene in passive guided devices and antennas from microwaves to THz have been by far less exploited. This is mainly because such structures require electric sizes of the order of the wavelength while the size of the first available graphene samples were much smaller. However, graphene chemical vapor deposition (CVD) now allows to obtain samples up to several centimeters, thereby increasing the research interest for the realization of passive devices at such frequencies [5-13].

Explanation on the unique band physics of graphene is beyond the scope of this document, but can be found in the plethora of references on the topic [1]. For our purpose, it is sufficient here to briefly discuss the main quantity of interest for graphene in electromagnetic (EM) applications. Since graphene is mono-atomic layer, it is perfectly modeled by surface conductivity tensor:

$$\bar{\bar{\sigma}} = \begin{pmatrix} \sigma_{xx} & \sigma_{xy} \\ \sigma_{yx} & \sigma_{yy} \end{pmatrix} = \begin{pmatrix} \sigma_d & \sigma_0 \\ -\sigma_0 & \sigma_d \end{pmatrix} \quad (1)$$

This conductivity tensor depends on graphene unique band structure and on a number of parameters including temperature, scattering rate, Fermi energy, electron velocity, pre-doping (not all these parameter are independent), as well as electrical and magnetic field bias. In practice other parameters such as defects in the potential polycrystalline nature of the graphene can further affect its properties. In general, the conductivity of graphene is very frequency-dependent, and can have completed different behavior e.g. at microwave and THz. In this context, it is pointless to try to numerically describe the conductivity behavior as a function of frequency and all different combinations of the above parameters. However, Fig. 1 provides an example of the graphene conductivity as a function of bias magnetic field. In general, the following main feature of graphene conductivity should be highlighted:

*Biasing and (an)isotropy:*

- An applied electric field bias injects more electron or holes carriers in graphene and thereby allows the dynamic control of both real and imaginary part of the conductivity.
- If a magnetic field bias is present or if spatial dispersion is considered, graphene is anisotropic. Therefore magnetic bias can be used to dynamically control graphene anisotropy, while graphene is mostly isotropic in the absence of magnetic bias.

*General isotropic conductivity behavior:*

- Though graphene is the best electrical conductor known, it is mono-atomic and thus the surface resistance is very high compared to metals at micro and mm-waves frequency, even with the possibility of doping and electric field biasing. In this frequency range graphene is thus mostly a moderate to bad conductive surface.

- Graphene plasma frequency is lower than for nobles metals, namely in the THz range. Therefore its conductivity has a totally different behavior at THz (equivalent to a negative real permittivity for a bulk material). Plasmonic propagation can thus be supported by graphene at THz, leading to extremely interesting properties for practical applications.

Finally, note that the conductivity of graphene has been directly or indirectly measured using different methods at DC up to THz (see e.g. [9-11]).

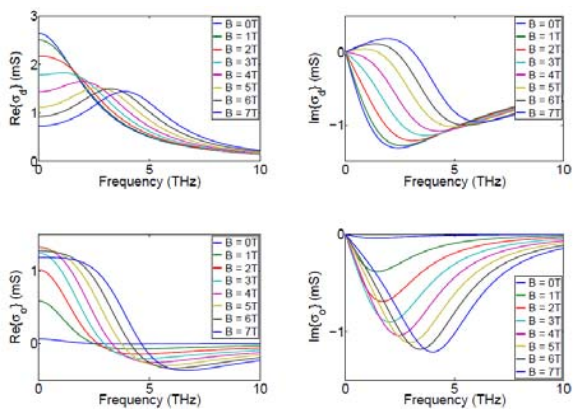

Fig. 1: The anisotropic conductivity of a monolayer graphene for different magnetic bias: (a) real part $\sigma_d$, (b) imaginary part of $\sigma_d$ (c) real part of $\sigma_0$ and (d) imaginary part of $\sigma_0$. Fermi energy= 0.34meV, scattering rate = 10 meV, velocity= 1.02e6 m/s.

II. GRAPHENE IN ANTENNA AND RELATED APPLICATIONS

This section lists various graphene devices being presently studied in our group, very briefly commenting on the motivation and main features in each category. We first discuss microwave and millimeter wave application, then exposing the most promising THz devices.

A. *Microwaves and millimeter-waves*

As discussed in the previous section, graphene is essentially a moderate to bad conductor at micro and mm-wave frequencies. However a few applications can still be envisioned in this frequency range, taking advantage of specific graphene properties.

*1) RF MicroelectroMechanical systems: RF-NEMS*

In a concept similar to RF-MEMS, a multilayer graphene (or graphite) membrane can be suspended over an air gap and mechanically actuated using an electrostatic force to create a RF switch or variable capacitor. The use of graphene here is mainly for integration and to benefit from the exceptional mechanical properties of graphene, achieving among other very low actuation voltage. Such devices will be fabricated in the near future, and Fig. 2 shows a schematic of such a graphene RF-NEMS as well as expected performance.

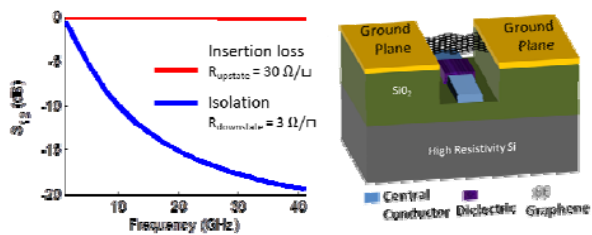

Fig. 2: Insertion loss and Isolation of a graphene based RF-NEMS capacitive shunt switch.

*2) Reconfigurable absorber*

Since graphene is quite resistive at micro and mm-waves, a straightforward application is to realize planar absorbers. Compared to other materials, the advantage is that –in addition to potential transparency and mechanical flexibility- such absorber can be dynamically controlled in a very simple and efficient manner. Fig. 3 shows an example of a multilayer graphene absorber and its performance under electronic control [14].

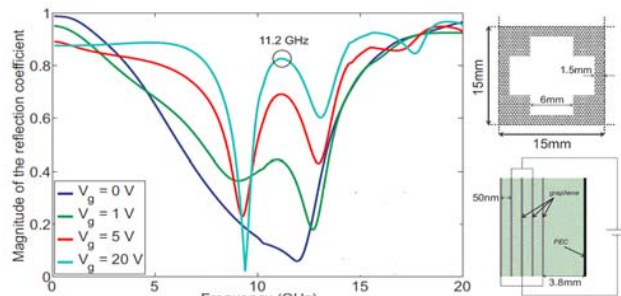

Fig. 3: Multilayer electronically-controllable graphene absorber [14].

*3) Graphene 'patch' antennas at microwave*

Though antennas with good efficiency cannot be realized at micro mm-wave using graphene, they might still be interesting for integration and transparency, in applications where low efficiency is tolerated (e.g. receiver in noise-limited regime). Such antennas were discussed in [13] and Fig. 4 highlights the potential for (limited) tuning but also the low efficiency of a graphene patch antenna at microwaves (graphene antennas at THz are discussed below).

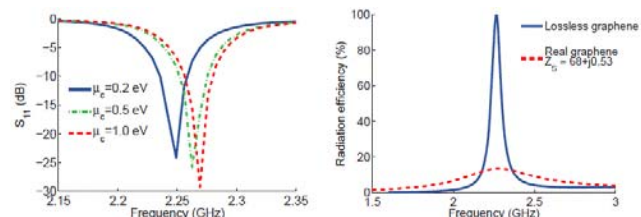

Fig. 4: Return loss and efficiency of a graphene patch at low microwave frequencies [13].

B. *THz and low-infrared*

As mentioned in section I, around its plasma frequency graphene conductivity becomes mostly reactive and can support plasmonic propagation or resonances. On the other hand, conventional technologies have significant limitations at THz frequencies and above (e.g. loss, possibility of dynamic control). Therefore we expect graphene to find its most interesting applications in this frequency range, and in the remainder of this section we list and briefly comment on some concrete applications presently under study in our group.

*1) Plasmonic TL and sensing*

It has been shown in different works already that the plasmonic mode supported by graphene are highly 'slow-wave', resulting in extreme device miniaturization, and are normally less lossy than using other materials. We have also recently shown that the properties of the plasmons can be beneficially affected not only by an electric bias but also by magnetic bias [15]. Moreover, the high confinement of the

TM surface wave field is extremely interesting for sensing applications, as had already been observed for noble metal plasmon at optical frequencies.

However a cause of growing concern in graphene plasmon modeling is spatial dispersion, which is usually neglected. We have demonstrated that the characteristics of surface plasmons strongly depend on the inherent spatial dispersion of the graphene sheet, even at the frequencies where intrabands contributions of graphene dominates, leading to waves with less mode confinement and higher losses than initially expected [17]. An example of the influence of spatial dispersion in the surface plasmons is shown in Fig. 5.

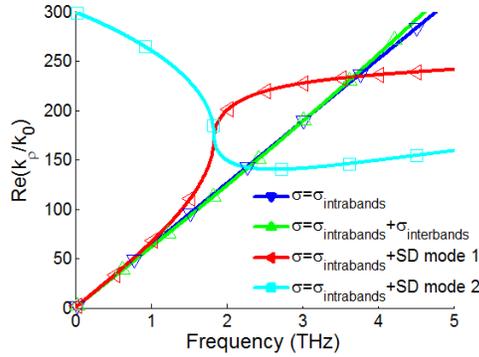

Fig. 5: Mode confinement of surfaces waves propagating along a spatially dispersive graphene sheet embedded into a homogeneous dielectric with $\varepsilon_r=6$. Reference results, computed without considering spatial dispersion, are included for comparison purposes. Graphene parameters are $\mu_c = 0$ eV, $\tau = 0.135$ ps and T = 300K.

*2) Plasmonic resonant antennas*

Graphene is a promising material for the realization of miniaturized resonant THz antennas. Notably, using simple dipole-like plasmonic resonant strips, good direct matching can be achieved to THz CW photomixers even with simple geometries [18]. The radiation efficiency is also good given the extremely small electrical size, leading to performances comparable to or better than metal implementations.

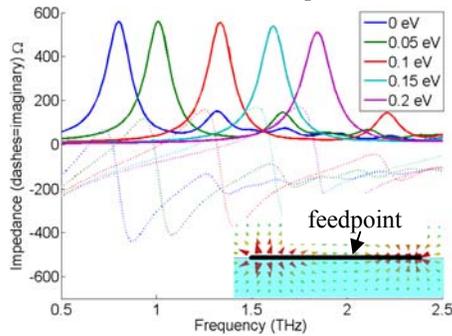

Fig. 6: Input impedance of the graphene plasmonic antenna for different values of the biasing voltage (expressed in terms of chemical potential). The inset shows the electric field around the antenna at the resonance.

Finally, here the electric field biasing of graphene is extremely interesting to provide easy and high-performance reconfiguration in the THz range. An illustration of a designed antenna performance is provided in Fig. 6. Note that [19] discussed the excitation of a resonance in a graphene patch under plane wave incidence, but no real feeding of the antenna was assessed, i.e. the graphene patch was just a scatterer there.

*3) Reflectarray and lens-array cells*

Reflectarray and lens-arrays are well-known antenna systems where an electromagnetic source illuminates an interface, generally made of sub-wavelength elementary 'cells', that imposes a *xy*-dependent phase shift in reflection or transmission, respectively. This non-uniform tailoring of phase shift along the surface allows to form a pseudo-arbitrary far field radiated beam as in conventional phased array. Note that a linear and phase-only version of this principle has also been recently reported at optical frequency, and was termed 'generalized Snell's law' [20]. Graphene is also extremely interesting for such applications, for the same reasons as the previous plasmonic THz examples (good plasmonic propagation, miniaturization, electronic control, etc). As an illustration Fig. 7 shows the reflection phase of a graphene patches ensemble designed for the aforementioned applications [21].

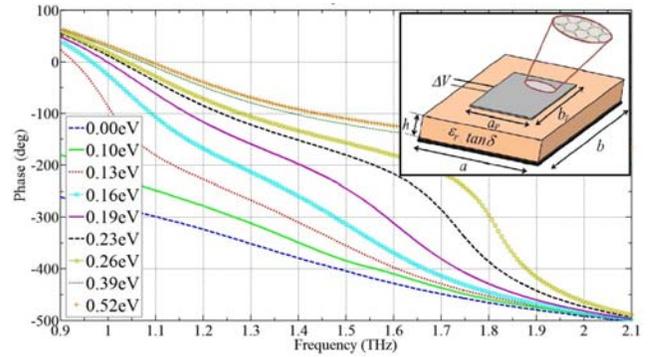

Fig. 7: Phase of the reflection coefficient for a 10μm-side square patch made of graphene for different values of the chemical potential ($\mu_c$), in the band from 0.9THz to 2.1THz [21].

*4) Frequency-selective surfaces and Polarizers*

A recent experimental demonstration of 2D periodic patterning of graphene in available in [22]. In [14], we proposed the design of different frequency-selective surfaces using graphene. One of the 2D periodic structure discussed there consist of two homogeneous graphene layers are etched periodically to create a lattice of slot (see Fig. 8). The slots are longer along *y* axis than *x* axis, which leads to different performance for *x* and *y* polarized waves. By proper adjustment of the dimensions, a tunable polarizer was designed, whose performance is shown in Fig. 8.

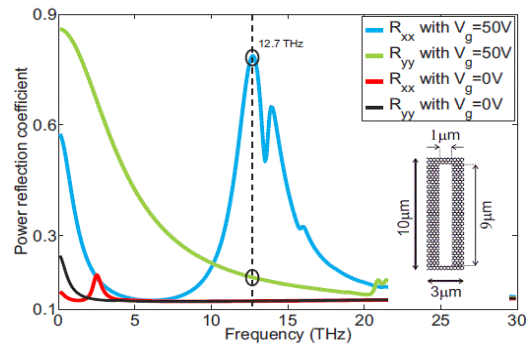

Fig. 8: Power reflection coefficients Rxx and Ryy versus frequency and bias voltage for a graphene reconfigurable polarizer consisting of two layers of graphene on a Si2N3 substrate (unit cell dimensions in figure inset, identical for each layer of graphene, themselves separated by 50nm).

*5) Giant Faraday rotation*

This last example is similar to the previous one in terms of periodic patterning a graphene surface, but uses graphene intrinsic anisotropy under magnetic bias field. It has been shown previously that a single graphene exhibits Giant Faraday rotation [10], a feature most interesting for various applications in optical telecommunications and laser technology (modulators, optical isolators and circulators), as well as for advanced magnetic field sensing. In [23], we demonstrate that this exceptional property can be manipulated through adequate nano-patterning, notably achieving giant rotation up to 6THz with features no smaller than 100nm. The effect of the periodic patterning on the Faraday rotation was predicted by a simple physical model, which was then verified and refined through accurate full-wave simulations. The resulting Faraday rotation angle is shown in Fig. 6.

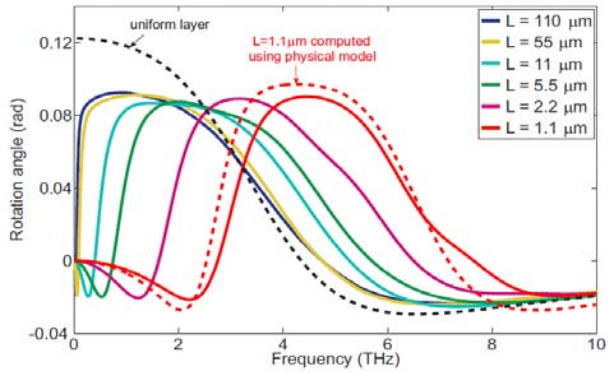

Fig. 9: Manipulation of the frequency of maximum faraday rotation of graphene via nano-patterning [21].

ACKNOWLEDGMENT

The author warmly thanks the different people involved directly or indirectly in the graphene electromagnetic effort at EPFL, Switzerland, including Dr. J.-S. Gomez-Diaz, M. Tamagnone, Dr. E. Carrasco, P. Sharma, C. Moldovan, and A. Ionescu, as well as Dr A. Fallahi from the Center for Free-Electron Laser Science, Hamburg, Germany.